\input{aipcheck}

\documentclass[
    ,final            
  ]
  {aipproc}

\layoutstyle{8x11single}

\begin{document}

\title{Modeling QCD for Hadron Physics}

\classification{24.85.+p, 12.38.Lg, 11.10.St, 14.40.-n}
\keywords      {Non-perturbative QCD, hadron physics, Dyson Schwinger equations, light quark mesons, pseuodoscalar decay constants, quark condensate as a property of the chiral pion wavefunction, deep inelastic scattering, valence quark parton distributions of pion and kaon.}

\author{P. C. Tandy}{
  address={Center for Nuclear Research, Department of Physics,  Kent State 
University, Kent OH 44242, USA}
}

\begin{abstract}
We review the approach to modeling  soft hadron physics observables based on the Dyson-Schwinger equations of QCD.    The focus is  on light quark mesons and in particular the pseudoscalar and vector ground states, their decays and electromagnetic couplings.   We detail the wide variety of observables that can be correlated by a  ladder-rainbow kernel with one infrared parameter fixed to the chiral quark condensate.   A recently proposed novel perspective in which the quark condensate is contained within hadrons and not the vacuum is mentioned.   The valence quark parton distributions, in the pion and kaon,  as measured in the Drell Yan process, are investigated with the same ladder-rainbow truncation of the Dyson-Schwinger and Bethe-Salpeter equations.
\end{abstract}

\maketitle


\section{DYSON--SCHWINGER EQUATIONS OF QCD}

A great deal of progress in the QCD modeling of hadron physics has been 
achieved through the use of the ladder-rainbow truncation of the Dyson-Schwinger
equations (DSEs).   The DSEs are the equations of motion of a
quantum field theory.  They form an infinite hierarchy of coupled
integral equations for the Green's functions ($n$-point functions) of
the theory.  Bound states (mesons, baryons) appear as poles in the appropriate
Green's functions, and, e.g., the Bethe-Salpeter bound state equation appears after taking residues in the DSE for the appropriate color singlet vertex.   The Faddeev equation for the 3-quark structure of baryons follows in a similar way from the residues of the appropriate vertex DSE.
For recent reviews on the DSEs and their use in hadron physics, see
Refs.~\cite{Roberts:1994dr,Tandy:1997qf,Alkofer:2000wg,Maris:2003vk}.   

In the Euclidean metric that we use throughout, the DSE for the dressed quark propagator is
\begin{eqnarray}
S(p)^{-1}  &=& Z_2 \, i\,/\!\!\!\!\!p + Z_4 \, m(\mu) + Z_1 \int^\Lambda_q \! g^2D_{\mu\nu}(p-q) \, 
        \frac{\lambda^i}{2}\gamma_\mu \, S(q) \, \Gamma^i_\nu(q,p)~,
\label{quarkdse}
\end{eqnarray}
where $D_{\mu\nu}(k)$ is the renormalized dressed-gluon propagator,
$\Gamma^i_\nu(q,p)$ is the renormalized dressed quark-gluon vertex.
We use $\int_q^\Lambda$ to denote $\int^\Lambda  d^4q/(2\pi)^4$ with $\Lambda$ being the
mass scale for translationally invariant regularization.    The renormalization condition is 
$S(p)^{-1}=i\gamma\cdot p+m(\mu)$ at a sufficiently large spacelike
$\mu^2$, with $m(\mu)$ the renormalized mass at renormalization scale $\mu$.
We use \mbox{$\mu=19\,{\rm GeV}$}.   The $Z_i(\mu, \Lambda)$ are renormalization constants.  

Bound state pole residues of the inhomogeneous Bethe-Salpeter equation (BSE) for the relevant vertex, yield the homogeneous BSE bound state equation 
\begin{eqnarray}
\Gamma^{a\bar{b}}(p_+,p_-) &=& \int^\Lambda_q \! K(p,q;P)S^a(q_+)
                                          \Gamma^{a\bar{b}}(q_+,q_-)S^b(q_-)~~,
\label{bse}
\end{eqnarray}
where $K$ is the renormalized $q\bar{q}$ scattering kernel that is
irreducible with respect to a pair of $q\bar{q}$ lines.  Quark
momenta are \mbox{$q_+ =$} \mbox{$q+\eta P$} and \mbox{$q_- =$} \mbox{$q -$}
\mbox{$(1-\eta) P$} where the choice of $\eta$ is equivalent to a definition of relative
momentum $q$; observables should not depend on $\eta$.  The meson momentum 
satisfies $P^2 = -M^2$.



A viable truncation of the infinite set of DSEs should respect
relevant (global) symmetries of QCD such as chiral symmetry, Lorentz
invariance, and renormalization group invariance.  For electromagnetic
interactions and Goldstone bosons we also need to respect color singlet  vector 
and axial vector current conservation.  The rainbow-ladder (LR) truncation achieves these ends by
the replacement
\mbox{$K(p,q;P)  \to -4\pi\,\alpha_{\rm eff}(k^2)\, D_{\mu\nu}^{\rm free}(k)
\textstyle{\frac{\lambda^i}{2}}\gamma_\mu \otimes \textstyle{\frac{\lambda^i}{2}}\gamma_\nu $}
along with the replacement of the DSE kernel  by
\mbox{$ Z_1 g^2 D_{\mu \nu}(k) \Gamma^i_\nu(q,p) \to 
 4\pi\,\alpha_{\rm eff}(k^2) \, D_{\mu\nu}^{\rm free}(k)\, \gamma_\nu
                                        \textstyle\frac{\lambda^i}{2} $}
where $k=p-q$, and $\alpha_{\rm eff}(k^2)$ is an effective running
coupling.   This truncation is the first term in a systematic
expansion~\cite{Bender:1996bb,Bhagwat:2004hn} of $K$; asymptotically, it reduces to leading-order perturbation theory.
These two truncations are mutually consistent: together they produce color singlet vector and axial-vector vertices satisfying their respective Ward identities.  This
ensures that the chiral limit ground state pseudoscalar bound states
are the massless Goldstone bosons from dynamical chiral symmetry
breaking~\cite{Maris:1997hd,Maris:1997tm}; and
ensures electromagnetic current conservation~\cite{Roberts:1996hh}.  
\begin{table}
\caption{DSE results~\protect\cite{Maris:1999nt, Maris:2000sk,Maris:1999bh,Jarecke:2002xd, Maris:2002mz,Ji:2001pj} for pseudoscalar and vector meson masses and electroweak decay constants, together with experimental data~\protect\cite{PDG04}.   Units are GeV except where indicated.   Quantities marked by $\dagger$ are fitted with the indicated current quark masses and the infrared strength parameter of the ladder-rainbow kernel.  \label{Table:model} }

\begin{tabular}{|l|cc|cc|cc|cc|cc|} \hline 
  \multicolumn{2}{|c|}{ }   & \multicolumn{3}{c|}{$m^{u=d}_{\mu=1 {\rm GeV}}$}  &  \multicolumn{3}{c|}{$m^{s}_{\mu=1 {\rm GeV}}$}  &   \multicolumn{3}{c|}{- $\langle \bar q q \rangle^0_{\mu=1 {\rm GeV}}$}    \\   \hline
 \multicolumn{2}{|c|}{ expt }   &  \multicolumn{3}{c|}{ 3 - 6 MeV}   &  \multicolumn{3}{c|}{  80 - 130 MeV } &  \multicolumn{3}{c|}{ (0.24 GeV)$^3$ }       \\     
\multicolumn{2}{|c|}{  calc }  &   \multicolumn{3}{c|}{ 5.5 MeV}  &  \multicolumn{3}{c|}{  125 MeV }  &   \multicolumn{3}{c|}{  (0.241 GeV)$^{3\dagger}$  }       \\ \hline
        & $m_\pi$ & $f_\pi$ & $m_K$ & $f_K$   &   $m_\rho$ &  $f_\rho$  & $m_K^\star$ & $f_K^\star$ &  $m_\phi$ &  $f_\phi$
 \\ \hline 
 expt  &   0.138  &  0.131  &   0.496   &   0.160 &   0.770  &  0.216  &   0.892 &  0.225   &   1.020   &  0.236    \\  
 calc  &   0.138$^\dagger$ &  0.131$^\dagger$ & 0.497$^\dagger$ &  0.155  &  0.742 & 0.207 & 0.936 & 0.241  &  1.072     &    0.259   \\ \hline  
\end{tabular}
\end{table}
\begin{table}
\begin{tabular}{|l|ccc|ccc|cc|} \hline
         & $r_\pi^2$ & ${r_K^+}^2$ & ${r_K^0}^2$ & $g^{\rho \pi \pi}$   &   $g_{\phi K K}$ &  $g_{K^\star K \pi}$  & $g_{\pi \gamma \gamma}$ & $r_{\pi \gamma \gamma}^2$      
  \\ \hline 
expt & 0.44~fm$^2$ & 0.34 & -0.054 & 6.02 & 4.64 & 4.60  &  0.50    &     0.42~fm$^2$    \\
calc & 0.45 & 0.38 & -0.086 & 5.4 & 4.3 & 4.1 &    0.50   &     0.41       \\ \hline     
  & $\frac{g_{\rho \pi \gamma}}{m_\rho}$ & $\frac{g_{\omega \pi \gamma}}{m_\omega}$ &  $\frac{g_{K^\star K \gamma}}{m_{K^+}}$ &  $\frac{g_{K^\star K \gamma}}{m_{K^-}}$  &  $\lambda_+(e3)$ & $\Gamma(K_{e3} )$  & $\Gamma(K_{\mu 3} )$  &            
 \\ \hline 
expt  & 0.74 & 2.31 & 0.83 & 1.28  &  0.028  &  7.6~10$^6~s^{-1}$  &  5.2   &    \\
calc  & 0.69 & 2.07 & 0.99 & 1.19  &  0.027   &    7.38       &      4.90  &      \\ \hline        
\end{tabular}
\end{table}
The ladder-rainbow kernel found to be successful for light quark hadrons~\cite{Maris:1997tm,Maris:1999nt} can be written
\mbox{$\alpha_{\rm eff}(k^2) =  \alpha^{\rm IR}(k^2) + \alpha^{\rm UV}(k^2) $}.
The IR term implements the strong infrared enhancement in the region
\mbox{$0 < k^2 < 1\,{\rm GeV}^2$} required for sufficient dynamical
chiral symmetry breaking.   The UV term preserves the one-loop renormalization group behavior of QCD: \mbox{$\alpha_{\rm eff}(k^2) \to \alpha_s(k^2)^{\rm 1 loop}(k^2)$} in the ultraviolet with 
\mbox{$N_f=4$} and \mbox{$\Lambda_{\rm QCD} = 0.234\,{\rm GeV}$}.   The strength of $\alpha^{\rm IR}$ along with two quark masses are fitted to $\langle\bar q q\rangle $, $m_{\pi/K}$.   

Selected light quark meson results are displayed in Table~\ref{Table:model}.   
The charge radii are obtained from the calculated charge form factors of $\pi$ and 
$K$~\cite{Maris:2000sk,Maris:1999bh} in the triangle diagram approximation.   This approximation is the dynamical partner of LR truncation:  charge is automatically conserved. Vector meson dominance is a natural outcome here since it is generated by the inhomogeneous BSE for the dressed photon-quark vertex.  Related to this, the decay constants for a vector meson to a pair of pseudoscalars are obtained by direct computation of the triangle diagram~\cite{Jarecke:2002xd}, or from analysis of the timelike electromagnetic form factors~\cite{Maris:2002mz}. 
The Abelian axial anomaly coupling $g_{\pi \gamma \gamma}$ and associated interaction radius are obtained from electromagnetic transition form factors~\cite{Maris:2002mz}, which also yield the radiative decay constants.    The three-body leptonic decays of the kaon~\cite{Ji:2001pj} are also  shown.  

It has remained true for many years that efforts~\cite{Alkofer:2002ne} to solve truncated versions of the DSEs for the QCD gauge sector n-point functions that determine the ladder-rainbow kernel have been unable to produce  sufficient kernel strength.   That is, an empirically acceptable amount of dynamical chiral symmetry breaking (and associated ground state hadron mass scales)  has long been underestimated that way.   Thus our infrared kernel component has remained semi-phenomenological while we seek to characterize it's deficiencies through application to new classes of observable.   Some efforts to go beyond the semi-phenomenological ladder-rainbow truncation so as to progressively reduce the amount of phenomenology have employed  an explicit 1-loop dressing of the gluon vertex:  a gluon loop~\cite{ Watson:2004kd}, or a pion loop~\cite{Fischer:2008wy}.    However, an amount of phenomenology is still needed in these approaches to reproduce light quark physics.    Furthermore,  a study~\cite{Matevosyan:2006bk,Matevosyan:2007cx}  of this type of a diagram-by-diagram  approach  to the vertex and BSE kernel, within a schematic model of simplified momentum dependence, shows that the produced color factors run rapidly out of control.   

 It is only recently that large volume lattice-QCD calculations~\cite{Cucchieri:2009zt} and analysis have established that the gluon 2-point function is non-zero in the IR limit, corresponding to a gluon effective mass of $\sim 0.5$~GeV, and that the ghost 2-point function is no more singular than its perturbative behavior.    These modern lattice-QCD results have recently~\cite{Aguilar:2010cn} been combined to produce a kernel for the quark DSE that yields an empirically acceptable amount of dynamical chiral symmetry breaking.    The gluon-quark vertex in  that approach might be used  in the recently developed method for generalized BSE kernel 
construction~\cite{Chang:2009zb}  to produce a physical and capable  pathway to a wider variety of hadron bound state calculations. 

\section{Novel Perspective on the Quark Condensate}

Recently it has been advocated that the quark condensate, rather than being a spacetime independent constant filling all space, is really a property of the Bethe--Salpeter wavefunction of hadrons~\cite{Brodsky:2010xf}.   Clear evidence for such a perspective is provided by the exact QCD expression  for the residue $\rho_\pi$ at the pseudoscalar meson pole in the 3-point function describing the coupling of a pseudoscalar field with quarks~\cite{Maris:1997hd}  
\begin{equation}
i\rho_\pi = -\langle 0 | \bar q i\gamma_5 q |\pi \rangle
 =  Z_4(\zeta,\Lambda)\; {\rm tr}_{\rm cd}
\int_{\Lambda_{\rm IR}}^\Lambda \frac{d^4 q}{(2\pi)^4}   
\;\gamma_5 \,S(q_+)\, \Gamma_\pi(q;P)\, S(q_-) \, .
\label{rhogen}
\end{equation}
Due to Goldsone's Theorem, and the resulting Goldberger-Treiman relation in which the chiral meson's invariant BSE amplitude associated with $\gamma_5$ is the chiral quark scalar self-energy divided by the chiral $f_\pi$, one can show rigorously that~\cite{Maris:1997hd}
\begin{equation}
\lim_{\hat m\to 0} \; f_\pi\; \langle 0 | \bar q \gamma_5 q |\pi \rangle
=   - Z_4(\zeta,\Lambda) \, {\rm tr}_{\rm cd}\int_{\Lambda_{\rm IR}}^\Lambda 
\frac{d^4 q}{(2\pi)^4} \ S_0(q;\zeta) 
 =   \langle \bar q q \rangle_\zeta^0\,.
\label{cond}
\end{equation}
Thus the chiral condensate is a property of the $q \bar q$ projection of the chiral pion Bethe-Salpeter wavefunction, and thus from this perspective is not defined away from this finite-sized bound state.    Although  it is true that the (point) pion cloud effect on the pion leads to an infinite charge radius, the $q \bar q$ dressed quark core remains of finite size and that is where 
Eq.~(\ref{cond}) places the condensate.     Furthermore this is in accord with quark and gluon confinement; these virtual fields only exist inside their hadronic containers. 

The quark condensate represents an enormous energy density, comparable to the interior of compact stellar objects.  The above perspective would eliminate the $10^{45}$ overestimate of the cosmological constant from the QCD vacuum energy density.    If the electroweak quanta turn out to be composites of higher mass entities, like technicolor quarks, the vacuum field we presently call the Higgs scalar is needed only inside those containers.   This could eliminate the $10^{56}$ discrepancy from that sector of the Standard Model.    

\section{Valence quark distributions in the pion and kaon}

Data for the momentum-fraction probability distributions of quarks and gluons in the pion have primarily been inferred from
Drell-Yan~\cite{Badier:1983mj,Betev:1985pg,Conway:1989fs} and direct photon production~\cite{Aurenche:1989sx} in pion-nucleon and pion-nucleus collisions, and semi-inclusive e$\,$p $\to$ e$\,NX$ reactions~\cite{Adloff:1998yg}.     For a recent review of nucleon and pion parton distributions 
see Ref.~\cite{Holt:2010vj}.
\begin{figure}[th] 
\hspace*{4mm}
\includegraphics[height=0.26\textheight]{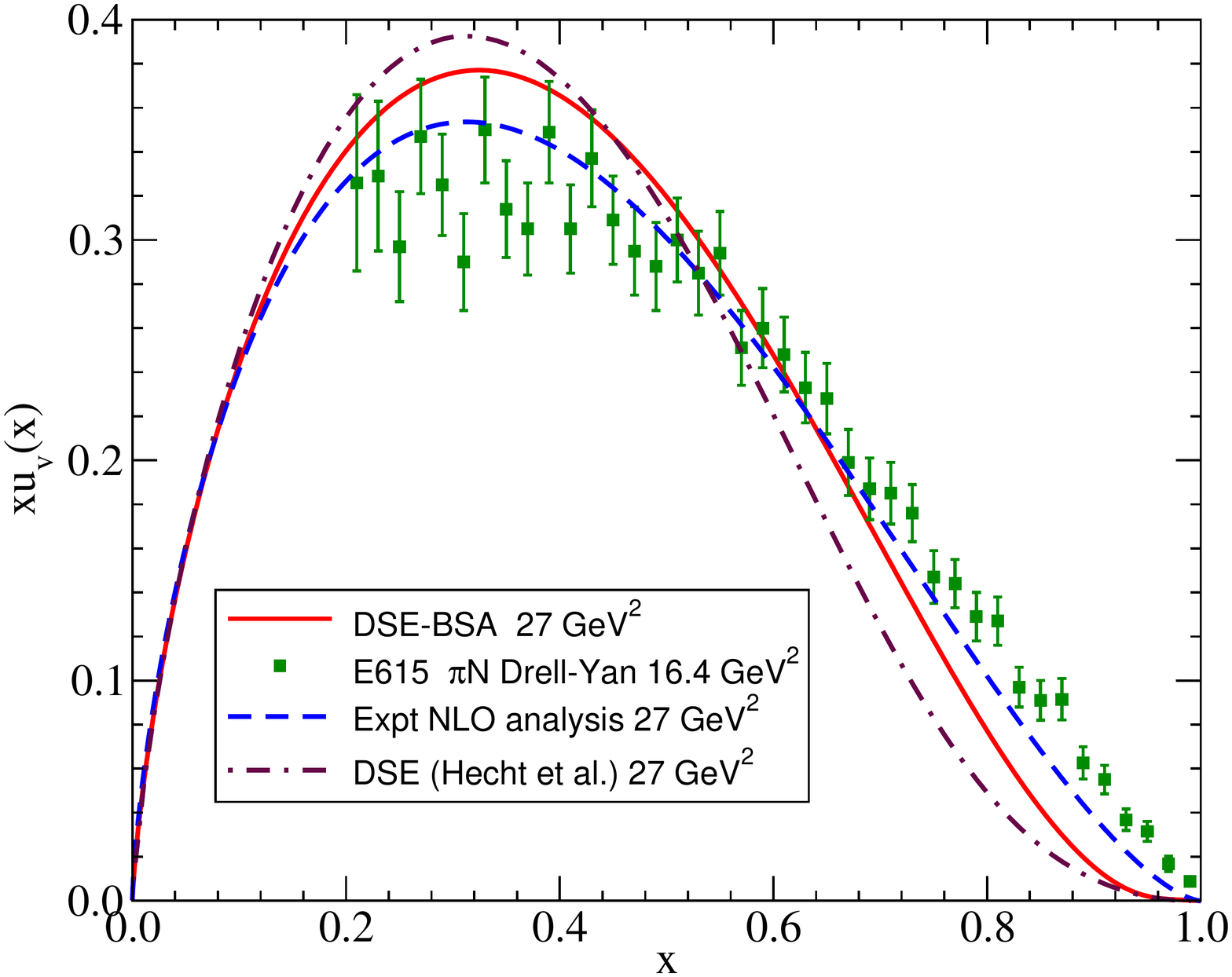}\hspace*{2mm}
\includegraphics[height=0.28\textheight]{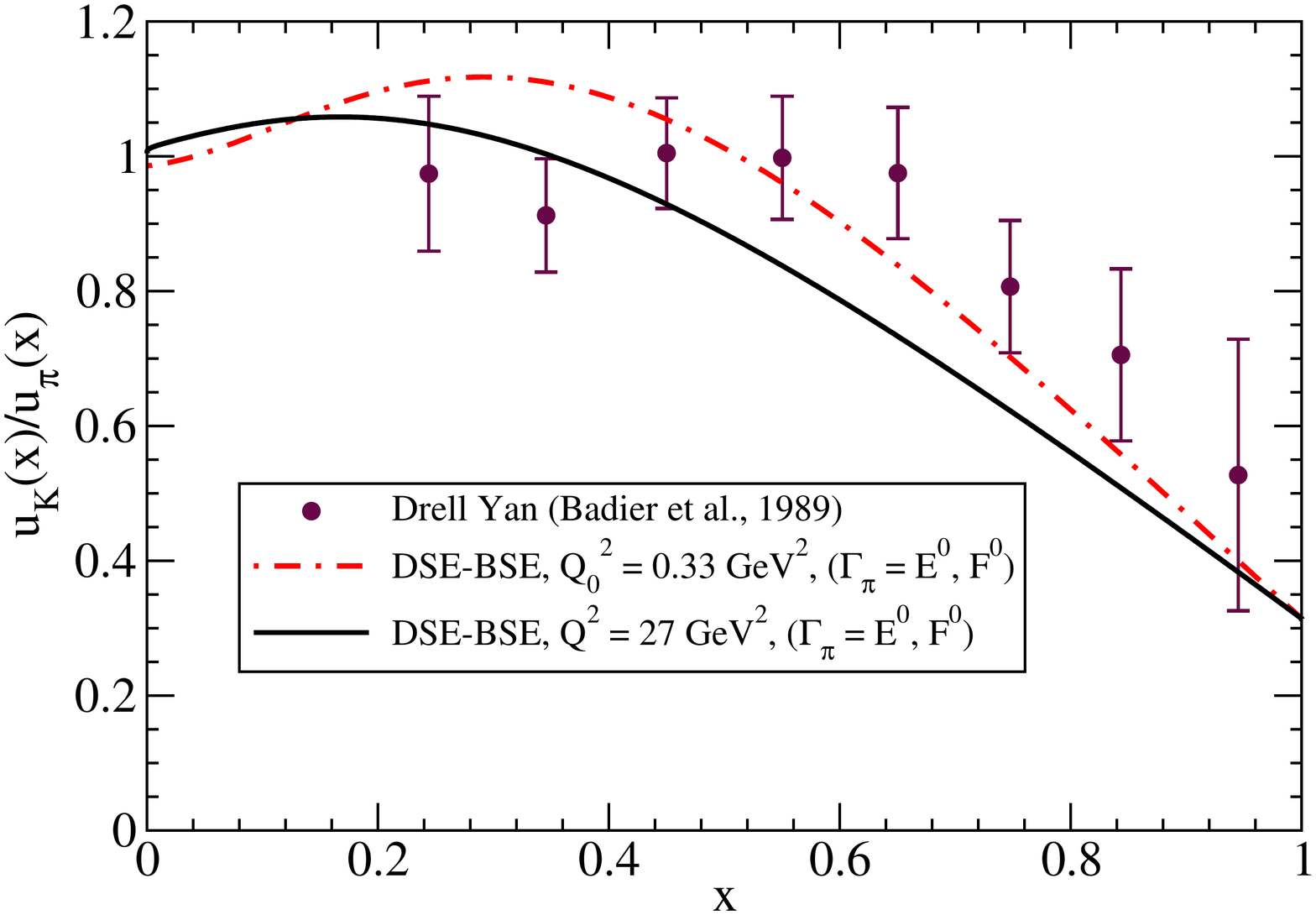}

\caption{{\it Left Panel}:    Pion valence quark distribution evolved to (5.2~GeV)$^2$.  Solid line is the full DSE-BSA calculation~\protect\cite{Nguyen_PhD10}; dot-dashed line is the semi-phenomenological DSE-based calculation of Hecht et al.~\protect\cite{Hecht:2000xa};   experimental data points are from~\protect\cite{Conway:1989fs} at scale (4.05~GeV)$^2$; the dashed line is the recent NLO re-analysis of the experimental data~\protect\cite{Wijesooriya:2005ir}.   {\it Right Panel}:   The ratio of u-quark distributions in the kaon and pion.   The solid line is our preliminary result from DSE-BSE 
calculations~\protect\cite{Nguyen:inprep10,Nguyen_PhD10,Holt:2010vj};  the experimental data is from~\protect\cite{Badier:1980jq,Badier:1983mj}.  \label{fig:pi_DSE+ratio} }
\end{figure} 

Lattice-QCD is restricted to  low moments of the distributions, not the
distributions themselves~\cite{Best:1997qp}.    Model calculations of deep inelastic scattering (DIS) parton distribution functions are challenging because it is necessary to have perturbative QCD features (including the evolution of scale)  coexisting with a covariant nonperturbative model made necessary by the bound state nature of the target.     Aspects of chiral symmetry have led to DIS calculations within the Nambu--Jona-Lasinio model~\cite{Shigetani:1993dx,Weigel:1999pc,Bentz:1999gx} but there are a number of difficulties with this approach~\cite{Weigel:1999pc,Bentz:1999gx}, among them a point structure for the pion BS amplitude at low model scale and a marked sensitivity to the regularization procedure due to the lack of renormalizability.    Constituent quark models have also been employed~\cite{Szczepaniak:1993uq,Frederico:1994dx}, with the difficulties encountered in such studies considered in Ref.~\cite{Shakin:1994rk}.   Instanton-liquid models~\cite{Dorokhov:2000gu} have also been used.    In these approaches, it is difficult to have pQCD elements join smoothly with nonperturbative aspects.
All of these issues can in principle be addressed if parton distribution functions can be obtained from a model based on the DSEs, and in particular styled after  the approach we have previously made for the related pion electromagnetic form factor~\cite{Maris:2000sk,Maris:1999bh}.    

In the Bjorken kinematic limit, DIS selects the most singular behavior of a correlator of quark fields of the target with light-like and causal distance separation $z^2 \sim 0^+ $.   With incident photon momentum along the negative 
3-axis, the kinematics selects \mbox{$z^+ \sim z_\perp \sim 0$}  leaving  $z^-$ as the finite distance conjugate to quark momentum component $xP^+$, where  \mbox{$ x = Q^2/2P \cdot q$} is the Bjorken variable, \mbox{$q^2 = -Q^2$} is the spacelike virtuality of the photon, and $P$ is the target momentum. To leading  order in the operator product expansion, the target structure functions are proportional to the probability distribution function  $q_f(x)$.   This in turn is given, in manifestly invariant form, by the  correlator~\cite{Jaffe:1983hp,Jaffe:1985je,Ellis:1991qj} 
\begin{equation}
q_f(x) = \frac{1}{4 \pi} \int d\lambda \, e^{-i x P\cdot n \lambda } 
\langle \pi(P) | \bar{\psi}_f (\lambda n) \, \gamma \cdot n  \, \psi_f(0) | \pi(P) \rangle_c ~~.
\label{Mink_dis_x_inv}
\end{equation}
where $f$ is a flavor label.    In the frame where the target has infinite momentum, $q_f(x)$  is the quantum mechanical probability that a single f-parton  has momentum fraction 
\mbox{$x=k\cdot n/P\cdot n $}~\cite{Ellis:1991qj}.   In the above $n^\mu$, and partner $p^\mu$, are  light-like  vectors satisfying \mbox{$n^2 = p^2 = 0$} and \mbox{$n \cdot p = 2$} and they form a convenient basis for the longitudinal sector of 4-vectors.   The dominant component of $q$ is parallel to $n$.   One can always choose a frame such that \mbox{$k \cdot n = k^+$} and \mbox{$k \cdot p = k^-$}.   Note that \mbox{$q_f(x) = - q_{\bar{f}}(-x) $}, and that the valence quark amplitude is \mbox{$ q^v_f(x) = q_f(x) - q_{\bar{f}}(x)$}.   It follows from Eq.~(\ref{Mink_dis_x_inv}) that 
\mbox{$ \int_0^1 dx \,  q^v_f(x) = $} \mbox{$ \langle \pi(P) | J_f(0)\cdot n | \pi(P) \rangle /2P\cdot n = F_\pi(0) = 1$}.   Approximate treatments should at least preserve  vector current conservation to automatically obtain the correct normalization for valence quark number.  


In ladder-rainbow truncation, which treats only the valence $q \bar q$ structure of the pion, Eq.~(\ref{Mink_dis_x_inv}) yields  the explicit form
\begin{equation}
q^v_f(x) = \frac{i}{2} \int_p^\Lambda {\rm tr}_{\rm cd} [ \Gamma_\pi(p,P) \, S(p)\, \Gamma^{(n)}(p;x) \, S(p) \,\Gamma_\pi(p,P)\, S(p-P) ]~~~,  
\label{Mink_dis_LR_Ward}
\end{equation}
where $\Gamma^{(n)}(p;x)$ is a generalization of the dressed vertex that a zero momentum photon has with a quark.  It satisfies the usual  inhomogeneous BSE integral equation  (here with a LR kernel) except that the inhomogeneous term is 
$\gamma \cdot n \, \delta(p\cdot n - xP\cdot n)$.   This selection of LR dynamics exactly parallels the symmetry-preserving dynamics of the corresponding treatment of the pion charge form factor at \mbox{$q^2 = 0 $} wherein the vector current is conserved by use of ladder dynamics at all three vertices and rainbow dynamics for all 3 quark propagators~\cite{Maris:2000sk,Maris:1999bh}.   Here the number of valence $u$-quarks (or $\bar d$) in the pion is automatically unity since the structure of Eq.~(\ref{Mink_dis_LR_Ward}), along with the canonical  normalization of the $q \bar q$ BS amplitude $\Gamma_\pi(p,P)$,  ensures \mbox{$ \int_0^1 dx \,  q^v_u(x) = 1$} because $ \int_0^1 dx \, \Gamma^{(n)}(p;x)$ gives the Ward Identity vertex.  

Eq.~(\ref{Mink_dis_LR_Ward}) is in Minkowski metric so as to satisfy the constraint on $p\cdot n$, but LR dynamical information on the various non-perturbative elements such as $S(p)$ and  $\Gamma_\pi(p,P)$ is available only in Euclidean metric~\cite{Maris:1999nt}.   Since $q_f(x)$ is obtained from the hadron tensor $W^{\mu \nu}$ which in turn can be formulated from the discontinuity  \mbox{$T^{\mu\nu}(\epsilon) -  T^{\mu\nu}(-\epsilon)$}, we observe that all enclosed singularities from the difference of Wick rotations cancel except for the cut that defines the object of interest.    With use of numerical solutions for dressed propagators  and BS amplitudes that give an accurate account of light quark hadrons, our DIS calculations  significantly extend  the exploratory study made in Ref.~\cite{Hecht:2000xa}.   That work employed phenomenological parameterizations of these elements.  

In Fig.~\ref{fig:pi_DSE+ratio} we display our DSE result for the valence $u$-quark distribution evolved to $Q^2 = (5.2~{\rm GeV})^2$ in comparison with $\pi N$ Drell-Yan data~\cite{Conway:1989fs} with a scale quoted as  $Q^2 = (4.05~{\rm GeV})^2$.   We also compare with  a recent NLO reanalysis of the data at scale $Q^2 = (5.2~{\rm GeV})^2$.   The distribution at the model scale $Q_0^2$ is evolved higher by leading order DGLAP.    The model scale is found to be \mbox{$Q_0 = 0.57 $}~GeV by matching the $x^n$ moments for $n=1,2,3$ to the experimental values given independently at (2~GeV)$^2$~\cite{Sutton:1991ay}.   
Our momentum sum rule result \mbox{$\int_0^1 dx\,  x\, (u_\alpha + \bar{d}_\alpha ) = $} \mbox{$0.74~(\alpha=\pi), 0.76~(\alpha=K)$} at $Q_0$ clearly shows the implicit inclusion of gluons as a dynamical entity  in a covariant approach.   
The large $x$ behavior in the form \mbox{$u(x) \sim (1-x)^\alpha$} has long provided a contrast between previous parameterizations of experiment (\mbox{$\alpha \sim 1.5  $})  and QCD-based approaches (\mbox{$\alpha = 2 +\gamma  $}) where $\gamma$ represents  anomalous dimensional behavior that increases with scale.    At our model scale, \mbox{$\alpha = 2.08$}.   Recently, the experimental data~\cite{Conway:1989fs} has been reanalyzed~\cite{Aicher:2010cb} by an accounting for soft gluon processes, and the resulting experimental $u(x)$ at  $Q^2 = (4~{\rm GeV})^2$, for \mbox{$x > 0.4$} is essentially identical to the 
Hecht et al. DSE calculation shown in Fig.~\ref{fig:pi_DSE+ratio}.  This eliminates the discrepancy
between QCD  and experiment here.  

The ratio $u_K/u_\pi$ measures the dynamical effect of the local environment.   In the kaon, the 
$u$-quark is partnered with a significantly heavier partner than in the pion and this shifts the probability to relatively lower $x$ in the kaon.   Our preliminary DSE model 
calculation~\cite{Nguyen:inprep10,Nguyen_PhD10,Holt:2010vj} is shown in Fig.~\ref{fig:pi_DSE+ratio}  along with available Drell Yan  data~\cite{Badier:1980jq,Badier:1983mj}.    Here we include only the leading two invariants of the pion BS amplitude, $E(k; P)$ and $F(k; P)$, where $k$ is $q \bar q$ relative momentum.   For both amplitudes only the zeroth Chebychev moment in $k\cdot P$ is employed .   This variable does not occur  in static quantum mechanics, nor in the Nambu--Jona-Lasinio point-coupling field theory model~\cite{Shigetani:1993dx} which also cannot accommodate the $k^2$ dependence.   We do not make such a point meson approximation here;  the $k^2$ dependence comes from the BSE solutions.   Nevertheless, the qualitative features of the ratio $u_K/u_\pi$  in Fig.~\ref{fig:pi_DSE+ratio} are adequately reproduced by a generalized Nambu--Jona-Lasinio model~\cite{Holt:2010vj}.

An explicit analysis of the exponent $\alpha$ in the large-$x$ behavior \mbox{$q(x) \sim (1-x)^\alpha  $} can be made in the limit where the propagators in Eq.~(\ref{Mink_dis_LR_Ward}) have the dynamical quark self-energy replaced by a constant (constituent) mass and the vertex  is taken to be $\gamma\cdot n \, \delta(p\cdot n - xP\cdot n)$.    These limits maintain consistency with the Ward Identity mentioned above.    We also truncate the pion Bethe-Salpeter vertex to just 
$\gamma_5 \, E_\pi( k^2) $ and take \mbox{$ E_\pi( k^2) = N_\pi/(  k^2 - \Lambda_\pi^2)  $} where 
the relative $q\bar q$ momentum is \mbox{$k = p - P/2  $}.  For \mbox{$x > 1/2$} the pole in the spectator quark propagator is the only one in the upper half plane and the $p^-$ integral may be easily evaluated by the residue theorem to yield
\begin{equation}
q^v(x) = \frac{4 N_c \pi^2}{(2\pi)^4} \int_0^\infty d^2p_\perp \; \frac{p_\perp^2 + m_q^2}{[x(1-x)P^2 -p^2_\perp - m_q^2]^2} \; E_\pi^2(k^2)~~~,
\label{NJL+E_perp_int}
\end{equation}
where now $k^2$ is evaluated at the $p^-$ pole.    It is convenient to change the integration variable to \mbox{$ \mu = - p^2 $} where the latter is the value at the $p^-$ pole. Then
\begin{equation}
q^v(x) = \frac{4 N_c \pi^2}{(2\pi)^4} \int_{\mu_m(x)}^\infty d\mu \; \frac{xP^2 + \mu + m_q^2}{[\mu + m_q^2]^2} \; E_\pi^2(-P^2/4 - (\mu + m_q^2)/2)~~~,
\label{NJL+mu_int}
\end{equation}
where \mbox{$\mu_m(x) = a/(1-x) - x m_\pi^2  $}, with \mbox{$a = x m_q^2  $}.   This divergence of the lower limit for large $x$ renders the result totally dependent upon the ultraviolet behavior of the propagators and bound state amplitudes.  

The leading large $x$ behavior of $q^v(x) \sim (1-x)^\alpha$ comes from the leading divergent term of each factor of the integrand.   The integral can be expressed
\begin{equation}
q^v(x) = N \int_0^\infty d\hat{\mu} \; \frac{\frac{a}{1-x} + b + \hat \mu}{[\frac{a}{1-x} + c + \hat \mu ]^2} \; (\frac{a}{1-x} + d + \hat \mu )^{-n}~~~,
\label{NJL+gen_int}
\end{equation}
where the physical case of  Bethe-Salpeter bound state amplitudes $E_\pi^2$ determined by one gluon exchange has \mbox{$ n = 2$}.    

The quantities $a, b, c, d $ depend on the mass-dimensioned quantities and scales in the system and have only a nonsingular dependence upon $x$.   Their details are not important.    A change of variable to \mbox{$ \bar \mu =  (1-x) \hat \mu /a$} shows that the integral scales as $[(1-x)/a ]^n$ when $a/(1-x)$ is greater than any physical mass scale in the system.     Note that it is the pair of bound state vertices that totally determine the large $x$ exponent:  if the argument of the bound state amplitude $E_\pi$ was not  singular at large $x$, the combined scaling effect of the propagators would vanish giving \mbox{$ \alpha = 0$}.    The presence of true bound state vertices provides a natural regularization of the integral and allows us to ignore explicit mathematical regularization at a higher scale.   In a renormalizable field theory such as QCD, there is an overall  regularization mass scale that can be taken to be much larger than any scale at which physical observable questions are to be asked of the formalism.  In the above case the renormalization scale is to be much larger than $a/(1-x)$ for values of $x$ where an asymptotic exponent is to be quoted.  

If one were to consider the limit of a Nambu--Jona-Lasinio model, the bound state vertices become constants and the model integral for $q^v(x) $ must be regularized, and the result, including the high $x$ exponent becomes dependent on regularization scheme~\cite{Holt:2010vj}.    However, independent of scheme, any thus obtained exponent \mbox{$ \alpha \neq 0$} would be unconstrained by the physical properties of the system.


\begin{theacknowledgments}
The authors would like to thank  C. D. Roberts, S. J. Brodsky and I. Cloet for helpful conversations
and suggestions.    This work has been partially supported by  the U.S. National Science Foundation under grant no. \ PHY-0903991, part of which constitutes  USA-Mexico collaboration funding in partnership with the Mexican agency CONACyT.
\end{theacknowledgments}



\bibliographystyle{aipproc}   

\bibliography{refsPM,refsPCT,refsCDR,refs,refsMAP}

\IfFileExists{\jobname.bbl}{}
 {\typeout{}
  \typeout{******************************************}
  \typeout{** Please run "bibtex \jobname" to optain}
  \typeout{** the bibliography and then re-run LaTeX}
  \typeout{** twice to fix the references!}
  \typeout{******************************************}
  \typeout{}
 }

\end{document}